\documentclass[aps,epsfig,floats,pre,twocolumn,showpacs]{revtex4-1}
\usepackage{graphicx}
\usepackage[latin2]{inputenc}
\usepackage{amsmath}
\usepackage{amssymb}
\usepackage{epsfig}
\usepackage{bm}
\usepackage{color}
\usepackage{stackengine}

\begin{document}

\title{Unraveling the structure of treelike networks from first-passage times 
of lazy random walkers}
\author{M. Reza Shaebani}
\altaffiliation{These authors contributed equally to this work.}

\author{Robin Jose}
\altaffiliation{These authors contributed equally to this work.}

\author{Christian Sand}
\altaffiliation{Current address: Heidelberger Institut f\"ur Theoretische 
Studien, 69118 Heidelberg, Germany\\
Corresponding author: shaebani@lusi.uni-sb.de
}

\author{Ludger Santen}

\affiliation{Department of Theoretical Physics $\&$ Center for Biophysics, Saarland 
University, 66123 Saarbr\"ucken, Germany}

\begin{abstract}
We study the problem of random search in finite networks with a tree topology, 
where it is expected that the distribution of the first-passage time $F(t)$ 
decays exponentially. We show that the slope $\alpha$ of the exponential tail 
is independent of the initial conditions of entering the tree in general, 
and scales exponentially or as a power law with the extent of the tree $L$, 
depending on the tendency $p$ to jump toward the target node. It is unfeasible 
to uniquely determine $L$ and $p$ from measuring $\alpha$ or the mean 
first-passage time (MFPT) of an ordinary diffusion along the tree. To 
unravel the structure, we consider lazy random walkers that take steps with 
probability $m$ when jumping on the nodes and return with probability $q$ 
from the leaves. By deriving an exact analytical expression for the MFPT 
of the intermittent random walk, we verify that the structural information 
of the tree can be uniquely extracted by measuring the MFPT for two randomly 
chosen types of tracer particles with distinct experimental parameters $m$ 
and $q$. We also address the applicability of our approach in the presence 
of disorder in the structure of the tree or statistical uncertainty in the 
experimental parameters.\\
\end{abstract}


%05.40.Fb  Random walks and Levy flights
%89.75.Fb	Structures and organization in complex systems
%89.75.Hc	Networks and genealogical trees
%02.50.Ey	Stochastic processes

\maketitle

\section{Introduction}
\label{Sec:Intro}
Diffusion and transport in complex environments are strongly influenced 
by the geometrical and topological properties of the underlying structures 
\cite{ben-Avraham00}. For example, the topology of the tree structure 
of human lung affects the absorption efficiency of diffusing oxygen 
\cite{Felici04}, the obstacle size and density determines the mean 
free path of light in turbid media \cite{Sadjadi}, the topology of 
Cayley trees influences the average displacement of quantum or 
classical random walkers \cite{Agliara08,Agliara14}, or the arrangement 
of magnetic bubbles in flashing potentials controls the anomalous 
behavior of the mean square displacement of paramagnetic colloidal 
particles \cite{Tierno16}.

\begin{figure}[b]
\centering
\includegraphics[width=0.47\textwidth]{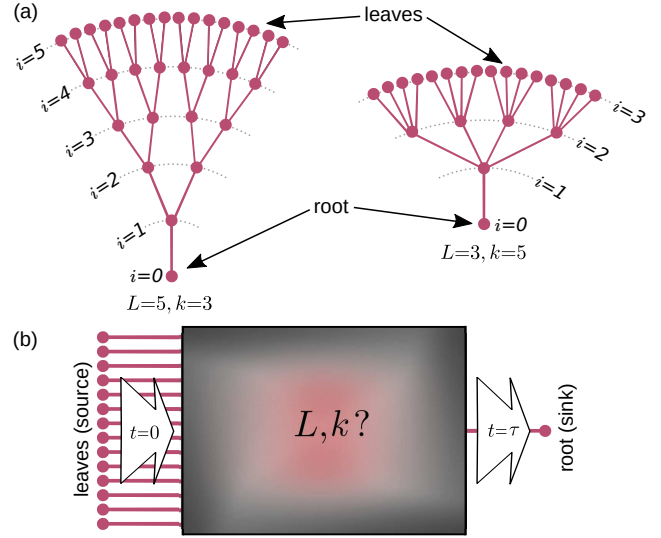}
\caption{(a) Examples of finite regular trees with the same number of leaves 
but different structure. (b) Schematic illustration of the measurement design. 
The tracer particle enters a tree, with unknown structural characteristics 
$L$ and $k$ (more generally $p$), from a leaf. It performs a random walk 
and eventually reaches the root after a first-passage time $\tau$. The mean 
first-passage time $\langle \tau \rangle$ is obtained by repeating the 
measurement for an ensemble of noninteracting tracer particles.}
\label{Fig1}
\end{figure}

Conversely, reconstructing the structure by means of the information 
obtained from the transport properties of tracer particles also constitutes 
an interesting subject. The idea of extracting the structural 
information of labyrinthine environments (or indirect evaluation of 
other quantities of interest in general) from the diffusional properties 
has attracted attention for a few decades \cite{Kac66,Mitra92,Mair99,
Krishna09,Chen12,Cooper16,Agliari07,Durian91,Maret97}: (i) It was 
suggested \cite{Kac66} that the geometry of the boundaries of a drum 
can be determined from the eigenvalues of the diffusion equation in a 
cage surrounded by absorbing walls; (ii) in porous structures, the 
porosity \cite{Mitra92}, surface-to-volume ratio of the voids 
\cite{Mair99}, degree of confinement and absorption strength 
\cite{Krishna09}, and permeability \cite{Chen12} were shown to be 
calculable from the diffusion propagator (more simply from the 
asymptotic diffusion coefficient in special cases); (iii) the 
temporal changes in the structure of foams and turbid media can 
be probed by the diffusive propagation of light \cite{Durian91,Maret97}; 
(iv) the geometrical properties of complex networks such as the 
number of triangles, loops, and subgraphs can be estimated from 
the first return time of random walks \cite{Cooper16}; and (v) 
as the last example, the time required for autocatalytic reactions 
on inhomogeneous substrates can be obtained from the mean time 
taken for the reactants to reach a reaction center or to encounter 
each other \cite{Agliari07}.

In this paper, we verify that the mean first-passage time (MFPT) 
of tracer particles to reach a target, as a conceptually simple and 
easily accessible transport quantity, can be employed to extract 
useful structural information. While we consider a treelike network 
in the present study, the idea can be extended to other complex 
networks and structures \cite{Sood07}. Branching morphologies constitute an important 
subset of complex structures, ranging from real systems (e.g.\ dendrimer 
macromolecules \cite{Helfand83,Wu12,Heijs04,Katsoulis02,Argyrakis00}, 
neuronal dendrites \cite{Spruston08,Hering01,Jose18}, and rivers 
\cite{Fleury01}) to virtual ones such as treelike graphs \cite{Szabo02,
Bollobas04}. To investigate diffusion on branched structures, they 
have been often modeled as regular treelike networks with, e.g., a 
given degree of the node $k$ representing the number of links connected 
to each node [see Fig.\,\ref{Fig1}(a)]. Well-studied examples include 
finite Cayley trees \cite{Helfand83,Wu12,Heijs04,Katsoulis02,Argyrakis00,
Redner01} and infinite Bethe lattices \cite{Hughes82,Monthus96,Cassi89}. 
A weight can be also assigned to each link \cite{Yook01,Almaas05}, as the 
real-world networks exhibit heterogeneity in the capacity of their links 
\cite{Barrat04,Krause03}. The advantage of regular trees is that the 
stochastic transport of particles along such structures can be mapped 
onto effective one-dimensional random walk models. By mapping Bethe 
lattices and Cayley trees onto 1D random walks, some basic quantities 
such as the mean square displacement, the probability of returning to 
the origin, and the first-passage times were calculated \cite{Helfand83,
Wu12,Heijs04,Katsoulis02,Argyrakis00,Hughes82,Monthus96,Cassi89}. For 
example, the MFPT to reach a target node in finite trees was shown to 
depend on the extent $L$ of the tree as well as the degree $k$ of the 
node (more generally on the probability $p$ to hop toward the target) 
\cite{Wu12,Heijs04,Khantha83,Skarpalezos13}. Thus, the structural 
parameters $L$ and $k$ [shown in Fig.\,\ref{Fig1}(a)] cannot be 
uniquely determined from the measurement of the MFPT of an ordinary 
random walk on the tree. The question arises of whether the prediction 
of the structural properties of trees from the MFPTs of other types of 
random walks is feasible. It is also not clear how far the possible 
predictions are robust in the presence of disorder in the extent of 
the tree, the degree of the nodes, or the capacity of the links.

To be able to unravel the structure, we increase the complexity of the 
dynamics of the tracer particles by introducing lazy random walkers that 
intermittently jump along the tree. We derive an exact analytical expression 
for the MFPT in terms of the waiting probabilities at nodes and dead 
ends (leaves), which enables us to uniquely determine the structure 
by measuring the MFPTs [see the schematic sketch in Fig.\,\ref{Fig1}(b)]. 
To identify the validity range of the theoretical results, we compare 
the analytical predictions to the simulation results in the presence of 
disorder in the structure of the tree or statistical uncertainty in the 
experimental parameters. Our results are also applicable to ordinary 
random walks along specific structures which induce temporal absorption 
along the path or at the dead ends. This is particularly relevant to 
transport in neuronal dendrites in the presence of biochemical cages 
along the dendritic tubes \cite{Spruston08,Hering01,Jose18}.

This paper is organized as follows. We first review biased random walks 
on bounded 1D domains and ordinary random walks on regularly branched 
trees in Sec.\,\ref{Sec:BRW}, and present an expression for the MFPT 
to travel from the leaves to the root of a finite tree. Next we introduce 
lazy random walkers in Sec.\,\ref{Sec:LRW} and demonstrate how the 
structure of the tree can be uniquely determined from the MFPTs of 
such tracer particles. In Sec.\,\ref{Sec:Disorder}, we show how the 
deviation of the MFPT from that of a regular tree enhances as the 
structural disorder or the experimental uncertainty grows. 
Sec.\,\ref{Sec:Conclusion} concludes the paper.

\section{First-passage times of biased random walks on finite 1D domains}
\label{Sec:BRW}

Stochastic motion of biased \cite{Garcia-Pelayo07,Pottier96,Weiss02,Benichou99,Mehra02} 
or persistent \cite{Garcia-Pelayo07,Pottier96,Weiss02,Shaebani14,Masoliver89} 
walkers in one dimension has been thoroughly investigated in the literature. 
Several aspects of 1D random walks, such as the influence of waiting \cite{Hafner16} 
or absorption \cite{Benichou99} along the path or at the boundaries \cite{Kantor07} 
on transport properties, has been studied. These studies also help understand 
the transport in other systems. For example, mapping of Bethe lattices and Cayley 
trees onto 1D random walks facilitates the calculation of the transport quantities 
of interest such as the first-passage times. While the MFPT to visit any 
specific target node on the Bethe lattice (i.e.\ an unlimited Cayley tree) is 
infinite, the MFPT in bounded domains such as Cayley trees is finite \cite{Wu12,
Heijs04,Khantha83,Skarpalezos13}. For instance, it was shown that the MFPT of 
traveling from the leaves to the root of a finite regular tree obeys the following 
relation \cite{Khantha83,Skarpalezos13}
\begin{equation}
\langle \tau \rangle = \displaystyle\frac{L}{2p-1} + 
\displaystyle\frac{1-p}{(2p-1)^2}\Big[\big(\frac{1-p}{p}\big)^L{-}1\Big],
\label{Eq:BRW}
\end{equation}
where $L$ denotes the extent of the tree and $p$ represents the tendency 
to hop toward the root at each node (corresponding to an effective 
bias toward the target in a 1D system). Here, the parameter $p$ equals to $p{=}
\frac1k$ in trees with uniform links, where $k$ is the coordination number 
(i.e.\ the degree of the nodes) as shown in Fig.\,\ref{Fig1}(a). More generaly, 
$p$ can represent the effective tendency to move toward the target including 
all the effects of branching at nodes, relative weights of the links, tendency 
to follow the shortest path toward the target, etc. The relevant systems are, 
e.g., weighted treelike networks \cite{Almaas05}, neuronal dendrites where 
the branches tend to taper toward the dead ends \cite{Jose18}, and stochastic 
packet transport in the Internet preferably along the shortest path 
\cite{Huisinga01,Kachhvah12} (though the latter structure contains loops). 
In deriving Eq.\,(\ref{Eq:BRW}), it was assumed that the reflection probability 
at the dead ends (leaves) equals the hopping probability toward the root at 
the bulk nodes \cite{Khantha83,Skarpalezos13}.

\begin{figure}[t]
\centering
\includegraphics[width=0.49\textwidth]{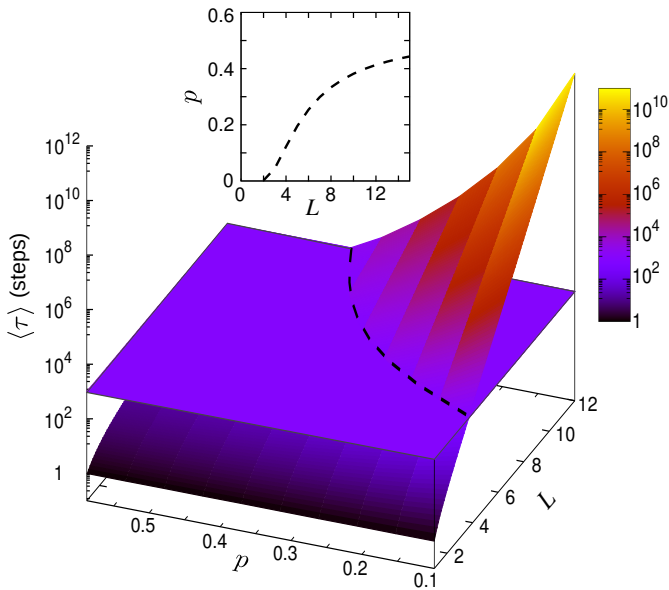}
\caption{Mean first-passage time $\langle \tau \rangle$ of traveling from the 
leaves to the root, obtained via Eq.\,(\ref{Eq:BRW}), versus the extent of the 
tree $L$ and the effective tendency $p$ to jump toward the target. The flat 
plane represents a constant MFPT, $\langle \tau \rangle{\simeq}1000\,
\text{steps}$, as an example of the measured MFPT in experiments. The dashed 
contour line marks the intersection of the two surfaces, i.e.\ the path along 
which the MFPT equals the measured value. Inset: The contour line in the 
$(L, p)$ phase space.}
\label{Fig2}
\end{figure}

According to Eq.\,(\ref{Eq:BRW}), the MFPT depends on both $L$ and $p$ parameters. 
Therefore, having access to the MFPT via the experiment illustrated in 
Fig.\,\ref{Fig1}(b) does not provide sufficient information to uniquely determine 
the structure, i.e.\ the tree extent $L$ and the node degree $k$ (or the 
effective upward tendency $p$). As shown in Fig.\,\ref{Fig2}, for a given 
measured value of the MFPT, one obtains a monotonic iso-MFPT contour line 
in the $(L, p)$ phase space as the set of possible solutions. 

\section{First-passage times of intermittent random walks}
\label{Sec:LRW}

We propose that the structural properties of an unknown tree can be deduced 
from the mean first-passage times of specific tracer particles with a tunable 
tendency to move along the tree. We consider random walks with waiting 
probabilities at nodes and leaves. When such lazy random walkers enter a 
tree from the leaves, explore the unknown structure, and exit from the root, 
the MFPT can be obtained in terms of the waiting probabilities. We show that 
having access to the MFPT for only two different random choices of the 
waiting probabilities enables us to uniquely determine the structure of 
the regular tree via the analytical framework developed in this section.

Let us consider the stochastic motion of an individual random walker on the 
nodes of a regular tree with the extent $L$. We can identify each 
node by its generation (i.e.\ its distance from the root) in regular trees. 
The generation of the nodes ranges from $0$ at the root to $L$ at the 
leaves. The number of nodes belonging to the same generation $i$ equals 
$k^{i{-}1}$ (for $i{>}0$), with $k$ being the coordination number of the 
nodes. Each tracer particle initially enters the tree from one of the leaves. 
When being on a bulk node, it either jumps to one of the neighboring nodes 
with probability $m$ or waits at its current node with probability $1{-}m$ 
at each time step. The dynamics is however different at the boundaries. 
At the leaves, the particle either returns to the interior of the tree 
with probability $q$ or waits with probability $1{-}q$. The other boundary, 
i.e.\ the root node with generation $i{=}0$, is treated as a trap. When the 
particle eventually reaches the root after a first-passage time $\tau$, it 
is not allowed to return to the tree (which corresponds to $m{=}0$ for the 
specific generation $i{=}0$).

In trees with uniform links and in the absence of other sources which induce 
preferency toward the target, $k$ determines the probability $\frac{m}{k}$ 
to jump to one of the neighboring nodes, thus, the relative tendency to move 
toward the target equals $p{=}\frac1k$. More generally, there can exist an 
additional net probability $m'$ to choose the root direction ($0{\leq}m'{
\leq}m$), induced by other possible effects (such as the hierarchical 
reduction of branch diameter toward the leaves or tendency to travel 
along the shortest path). In such a case, the total probability to jump 
from a bulk node toward the root or each of the leaves is $\frac{m{-}m'}{k}
{+}m'$ or $\frac{m{-}m'}{k}$, respectively. Therefore, the relative tendency 
to move toward the root effectively equals $p{=}\frac{m{-}m'}{m}\frac1k{+}
\frac{m'}{m}$ and the tendency to hop to any of the nodes toward the leaves 
is $\frac{m{-}m'}{m}\frac1k$. 

\begin{figure}[t]
\centering
\includegraphics[width=0.47\textwidth]{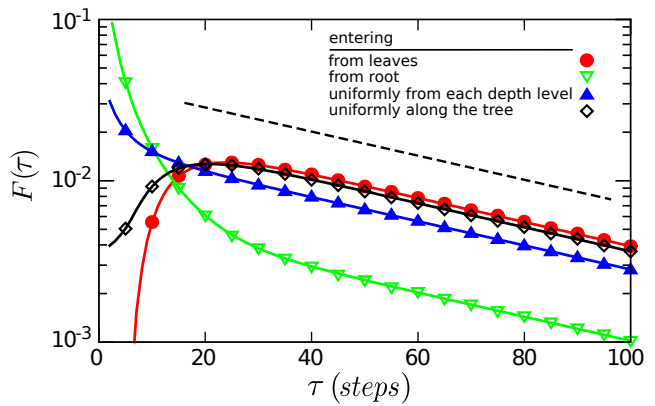}
\caption{First-passage time distribution for $L{=}6$ and 
$p=m=q=0.5$ and for different initial conditions of 
entering the tree. The solid lines are the analytical 
prediction of Eq.\,(\ref{Eq:Fz}) and the symbols denote 
the simulation results. The exponential dashed line is 
given by Eq.\,(\ref{Eq:Tail}).}
\label{Fig3}
\end{figure}

\begin{figure*}[t]
\centering
\includegraphics[width=0.97\textwidth]{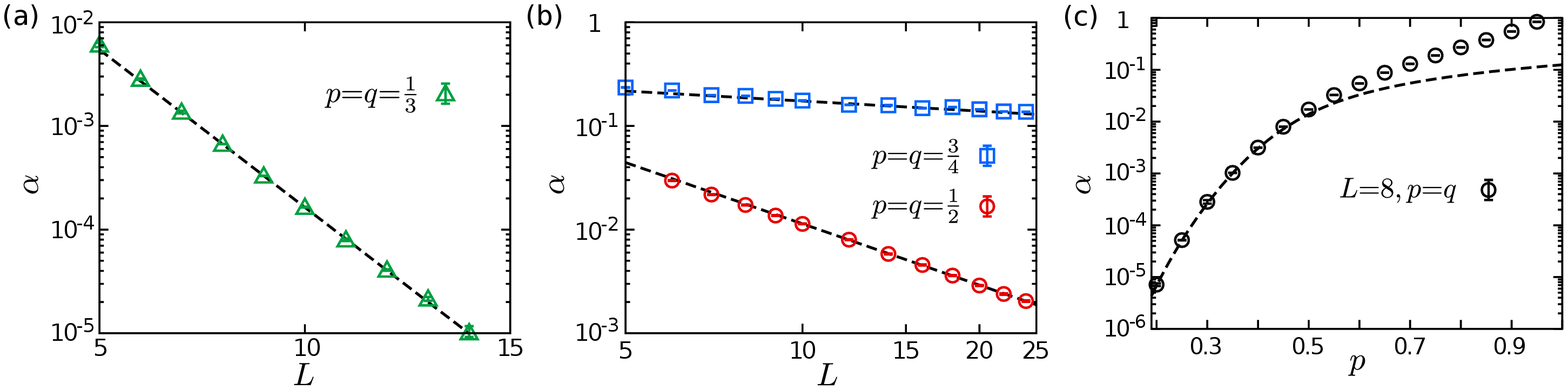}
\caption{The slope $\alpha$ of the exponential tail versus the 
extent of the tree for (a) $p{<}\frac12$ (log-lin scales) and 
(b) $p{\geq}\frac12$ (log-log scales) at $m{=}1$. The dashed 
lines represent exponential or power-law fits. (c) $\alpha$ 
versus $p$ at $L{=}8$ and $m{=}1$. The dashed line represents 
$\alpha{=}1/\langle \tau \rangle$ for the MFPTs obtained for 
different values of $p$ at $L{=}8$, $m{=}1$, and $q{=}p$.}
\label{Fig4}
\end{figure*}

In the following, we solve the problem of the first-passage time 
from leaves to root for the set of parameters $\{L,p,m,q\}$. 
However, one can straightforwardly follow the proposed approach 
to calculate the MFPT between two arbitrary generations of the 
tree. By introducing the probability distribution $P\!\!_{_i}(t)$ 
of being on a node with generation $i$ at time step $t$, we 
practically map the problem onto an effective biased random 
walk on a 1D domain in the presence of temporal absorption along 
the path and at one of the boundaries. Using the initial condition 
$P\!\!_{_i}(0){=}\delta_{i,L}$, we construct a set of master 
equations for the dynamical evolution of $P\!\!_{_i}(t)$. The 
probability evolves at the root ($i{=}0$) and leaves ($i{=}L$) 
as $P\!\!_{_0}(t){=}P\!\!_{_0}(t{-}1) {+} m \, p \, P\!\!_{_1}(t
{-}1)$ and $P\!\!_{_L}(t){=} m \, (1{-}p) \, P\!\!_{_{L{-}1}}(t
{-}1) {+} (1{-}q) \, P\!\!_{_L}(t{-}1) {+} \delta_{0,t}$. At a 
bulk node with generation $i$, the evolution of $P\!\!_{_i}(t)$ 
follows $P\!\!_{_i}(t) {=} m \, (1{-}p) \, P\!\!_{_{i{-}1}}(t
{-}1) {+} (1{-}m) \, P\!\!_{_i}(t{-}1) {+} m \, p \, P\!\!_{_{i
{+}1}}(t{-}1)$. By defining the $z$ transform $P\!\!_{_i}(z){=}
\sum\limits_{t{=}0}^{\infty} P\!\!_{_i}(t)\,z^t$, we obtain the 
following set of coupled equations 
\begin{equation}
\left\{
\begin{array}{ll}
P\!\!_{_0}(z) &\!\!\!= z \, P\!\!_{_0}\!(z) {+} 
m \, p \,z \, P\!\!_{_1}\!(z),\vspace{1mm}\\
P\!\!_{_1}(z) &\!\!\!= (1{-}m) z \, P\!\!_{_1}\!(z) {+} 
m \, p \, z \, P\!\!_{_2}\!(z),\\
\;\;\;\;\Shortstack{ . . .} & \\
P\!\!_{_i}(z) &\!\!\!= m (1{-}p) z \, P\!\!_{_{i{-}1}}
\!(z) {+} (1{-}m) z \, P\!\!_{_i}(z)  \\
&\hspace{26.2mm}{+} m \, p \, z \, P\!\!_{_{i{+}1}}\!(z),\\
\;\;\;\;\Shortstack{ . . .} & \\
P\!\!_{_{L{-}1}}\!(z) &\!\!\!= m (1{-}p) z \, P\!\!_{_{L
{-}2}}\!(z) {+} 
(1{-}m) z \, P\!\!_{_{L{-}1}}\!(z) {+} q \, z \, P\!\!_{_L}
\!(z),\vspace{1mm}\\
P\!\!_{_L}\!(z) &\!\!\!= m (1{-}p) z \, P\!\!_{_{L{-}1}}
\!(z) {+} (1{-}q) z \, P\!\!_{_L}\!(z) {+} 1.
\end{array}
\right.
\label{Eq:MasterEqs}
\end{equation}
After some algebra we obtain $P\!\!_{_i}(z)$ in the general 
form. Then, the $z$ transform of the FPT distribution to reach 
the root can be evaluated as $F(z){=}m\,p\,z\,P\!\!_{_1}(z)$ 
\cite{Redner01}. We derive an exact expression for the $z$ 
transform of the FPT distribution \cite{Jose18} 
\begin{equation}
F(z) {=} \displaystyle \frac{2^{L+1} \, q \, A(z,m,p)}{\big(
H_+^L {-} H_-^L\big) B(z,q,m,p) {+} \big(H_+^L {+} H_-^L\big) 
q \, A(z,m,p)},\vspace{3mm}
\label{Eq:Fz}
\end{equation}
with $A(z,m,p){=}\displaystyle\sqrt{1{+}2(m{-}1)z{+}
\big[1{-}2m{+}(1{-}2p)^2\,m^2\big]z^2}$, $B(z,q,m,p)\!=\!
q\big(1{-}(m{-}1)z\big)+p\,m({-}2{+}2\,z {-}q\,z)$, and 
$H_{\pm}{=}\displaystyle\frac{1}{mpz}\big[1{+}(m{-}1)z{\pm}
A(z,m,p)\big]$. The first-passage time distribution $F(\tau)$ 
can be obtained by inverse $z$ transforming of $F(z)$. In order 
to check our lengthy analytical results for correctness, we 
compare them with the results of Monte Carlo simulations in 
Fig.\,\ref{Fig3} and find them in perfect agreement.

Fig.\,\ref{Fig3} shows that the tail of $F(\tau)$ decays 
exponentially with a slope $\alpha$, which is independent of 
the initial conditions of entering the tree. Expectedly, 
$\alpha$ varies with the structural properties $L$ and $p$ 
as well as the waiting probabilities $m$ and $q$. While the 
tail behavior of the lengthy expression $F(\tau)$ cannot be 
necessarily expressed in a closed form in general, at least 
the existence of an exponential tail can be proven. $F(z)$ 
can be represented as the inverse of a polynomial $g(z,p,m,q,L)$ 
with $j$ roots ($j{\leq}L$), thus, can be written as 
$F(z){=}\displaystyle\frac{1}{(1{-}a_{_1} z)^{b\!_{_1}} 
\cdot \cdot \cdot (1{-}a_{_j} z)^{b\!_{_j}}}$, where the 
prefactors of $z$ are functions of the parameter set 
$\{L,p,m,q\}$ and $b\!_{_1}{,...},b\!_{_j}{\leq}L$. As a 
result, $F(\tau)$ can be written as a sum of $a_{_k}^t$ 
terms by partial fraction decomposition of $F(z)$ and applying 
inverse $z$ transform. Therefore, one can approximate $F(\tau)$ 
by the leading exponential term $a_{_{k,\text{max}}}^t$ 
in the long-time limit. In view of the difficulty to extract 
the roots of the polynomial and deduce a general form for 
the exponential asymptotic scaling, we choose the given 
set of parameter values in Fig.\,\ref{Fig3} and reconstruct 
the master equations (\ref{Eq:MasterEqs}) for the four 
different initial conditions of entering the tree introduced 
in the figure. It can be shown that the leading term of 
$F(\tau)$ follows 
\begin{equation}
F(\tau)\sim \exp\Big[-\ln\!\Big(\frac{4}{2{+}\sqrt{2{+}
\sqrt{3}}}\Big)\;t\Big]
\label{Eq:Tail}
\end{equation}
for all the initial conditions. Thus, the slope $\alpha$ 
of the exponential tail can be deduced as $\alpha{=}\ln\!
\Big(4/\big(2{+}\sqrt{2{+}\sqrt{3}}\big)\Big)$ for the given 
set of parameters in Fig.\,\ref{Fig3}. 

\begin{figure*}[t]
\centering
\includegraphics[width=0.97\textwidth]{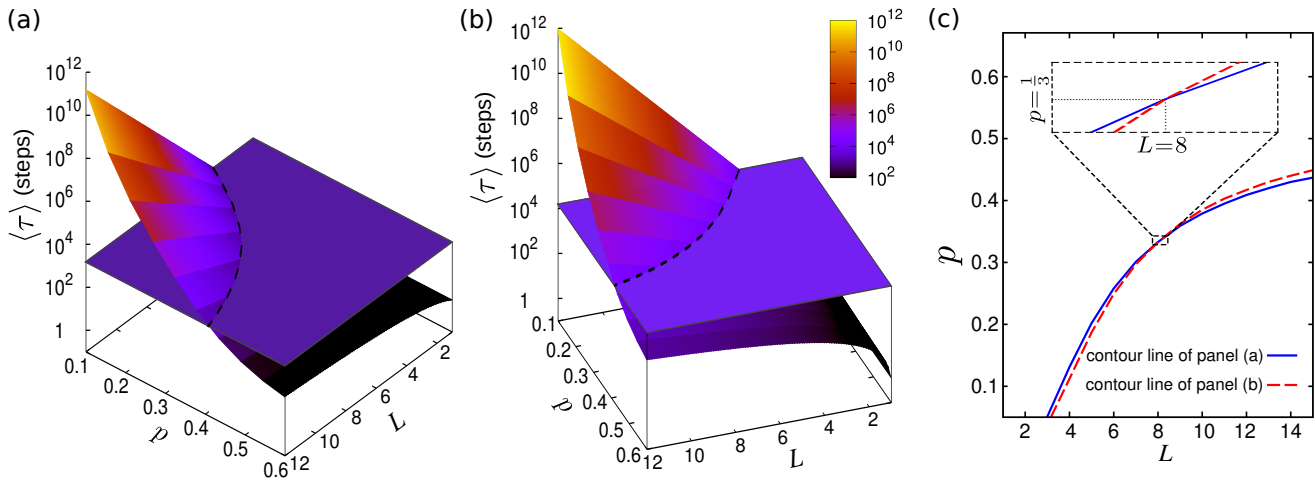}
\caption{Mean first-passage time via Eq.\,(\ref{Eq:MFPT}) versus 
the effective tendency $p$ to jump toward the target and the tree 
extent $L$ at (a) $m{=}1$, $q{=}\frac13$ and (b) $m{=}0.05$, 
$q{=}0.8$. The flat (constant MFPT) planes correspond to the MFPTs 
measured by tracer particles in experiments (chosen to be $\langle 
\tau \rangle {\simeq}1500$ and $15000$ steps in panels (a) and (b), 
respectively). The dashed line marks the intersection of the two 
surfaces, i.e.\ the contour line along which the MFPT equals the 
measured value. (c) Intersected contour lines in the $(L, p)$ 
phase space. Inset: A zoomed view of the same plot near the 
intersection, which determines the unknown values of the 
structural parameters.}
\label{Fig5}
\end{figure*}

By extracting $\alpha$ for other values of the structural 
parameters $L$ and $p$, we find that $\alpha$ scales 
exponentially with the extent of the tree $L$ if $p{<}
\frac12$. However, a crossover to power-law scaling 
occurs for $p{\geq}\frac12$, as shown in Figs.\,\ref{Fig4}(a),(b).  
The overall shape of $F(\tau)$ exhibits a plateau or even 
develops a peak at short times in general. The characteristic 
time to converge to the exponential tail behavior reduces 
with decreasing $p$, such that the entire distribution $F(\tau)$ 
follows an exponential form in the limit $p{\rightarrow}0$. 
Consequently, the mean value of $F(\tau)$ (i.e.\ the MFPT 
$\langle\tau\rangle$) is inversely related to the slope 
$\alpha$ at small values of $p$, as expected for exponential 
distributions. With increasing $p$, the form of $F(\tau)$ 
changes, thus, the deviations from $\alpha{=}\frac{1}{\langle
\tau\rangle}$ grow, as shown in Fig.\,\ref{Fig4}(c).  

The MFPT can be calculated as $\langle\tau\rangle = z 
\frac{d}{d z}F(z)\Bigr|_{z{\rightarrow}1}$. We expand 
Eq.\,(\ref{Eq:Fz}) around $z{=}1$ up to first order 
terms, as $F(z){\sim}F(z)\Bigr|_{z{\rightarrow}1}{+}(z{-}1)
\frac{d}{d z}F(z)\Bigr|_{z{\rightarrow}1}{+}\mathcal{O}
\Big((z{-}1)^2\Big)$, and obtain the following exact 
expression for the MFPT
\begin{equation}
\langle \tau \rangle {=}
\begin{cases} 
\displaystyle \frac{(m{-}q)L{+}q\,L^2}{m\,q},  
& p{=}\frac12, \vspace{1mm}\\
\displaystyle\frac{L}{m\,(2p{-}1)} {+} 
\frac{p\,q{-}p\,m\,(2p{-}1)}{m\,q\,(2p{-}1)^2}\Big((
\frac{1}{p}{-}1)\!^{^L}\!\!{-}1\Big), & p{\neq}\frac12. 
\end{cases}
\label{Eq:MFPT}
\end{equation}
The MFPT diverges in the limit $L{\rightarrow} \infty$ as expected 
for infinite structures. It can be also seen that the MFPT scales 
linearly (exponentially) with the extent of the tree $L$ in the 
limit $p{\rightarrow}1$ ($p{\rightarrow}0$), as the exponential 
term on the right-hand side of Eq.\,(\ref{Eq:MFPT}) vanishes 
(dominates). In the absence of waiting ($m{=}1$) and assuming 
that the reflection probability $q$ at the leaves equals the 
hopping probability $p$ toward the root at the bulk nodes 
($q{=}p$), Eq.\,(\ref{Eq:MFPT}) for $p{\neq}\frac12$ reduces 
to Eq.\,(\ref{Eq:BRW}). It is also notable that the total number 
of nodes $N$ in a regular tree is given as $N{=}2^L$. Thus, the 
first (last) term of $\langle \tau \rangle$ for $p{\neq}\frac12$ 
grows logarithmically (linearly) with the number of nodes \cite{Wu12}.

According to Eq.\,(\ref{Eq:MFPT}), the MFPT of an intermittent 
random walk to travel from the leaves to the root depends on the 
set of parameters $\{L,p,m,q\}$. Let us suppose that the intrinsic 
dynamics of the lazy random walkers, characterized by $m$ and 
$q$ parameters, can be tuned before they start to explore the 
structure as tracer particles. For a given set of $m$ and $q$, 
one obtains a surface in the $(L, p, \langle \tau \rangle)$ 
phase space via Eq.\,(\ref{Eq:MFPT}). The intersection of this 
surface with a flat plane representing the measured MFPT results 
in an iso-MFPT contour line in the $(L, p)$ plane. An example 
is presented in Fig.\,\ref{Fig5}(a) for $m{=}1$ and $q{=}\frac13$, 
assuming that the measured MFPT is $\langle \tau \rangle{\simeq}
1500$ steps. By repeating this procedure for a different set of 
$m$ and $q$ (with a different $\frac{m}{q}$ ratio), we can obtain 
another contour line in the $(L, p)$ plane which intersects the 
previous one and, thus, uniquely determines the structure (i.e.\ 
$L$ and $p$ parameters). Fig.\,\ref{Fig5}(b) shows an example, 
where new parameter values $m{=}0.05$ and $q{=}0.8$ are chosen. 
Assuming that the measured MFPT for this lazy random walker is 
$\langle \tau \rangle{\simeq}15000$ steps, we obtain the second 
contour line. The intersected contour lines in Fig.\,\ref{Fig5}(c) 
reveal that the unknown tree has the extent $L{=}8$ and the effective 
tendency $p{\simeq}\frac13$ to move toward the root (corresponding 
to the node degree $k{=}3$ in the case of uniform links). Therefore, 
having access to the MFPTs of lazy random walkers provides sufficient 
information to unravel the structure.

\begin{figure}[t]
\centering
\includegraphics[width=0.43\textwidth]{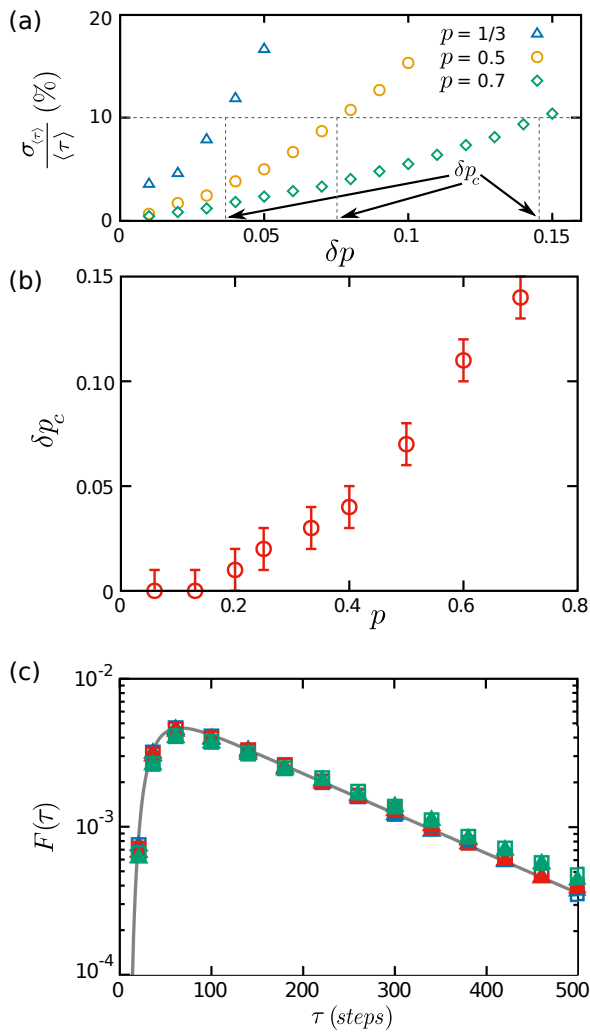}
\caption{(a) The standard deviation of the MFPT in simulations, 
scaled by the analytical prediction for $\langle \tau \rangle$ 
via Eq.\,(\ref{Eq:MFPT}), in terms of the variation range $\delta 
p$ around its mean value $p$ at $L{=}10$ 
and $m{=}q{=}1$. The results are shown for three different values 
of $p$. The horizontal dashed line corresponds to $10\%$ error 
in the measured MFPT in simulations compared to the analytical 
prediction. (b) The critical variation range of the effective 
tendency $\delta p_c$ (which causes $10\%$ differences between 
simulation and theory) in terms of $p$. (c) The FPT distribution 
obtained via Eq.\,(\ref{Eq:Fz}) (solid line) or simulations with 
$\delta p{\sim}0.02$ (blue symbols), $0.04$ (red symbols), and 
$0.10$ (green symbols). The results of uniform probability 
distributions of $p$ (full triangles) are compared to those 
obtained from normal distributions (open squares)}.
\label{Fig6}
\end{figure}

\begin{figure}[t]
\centering
\includegraphics[width=0.47\textwidth]{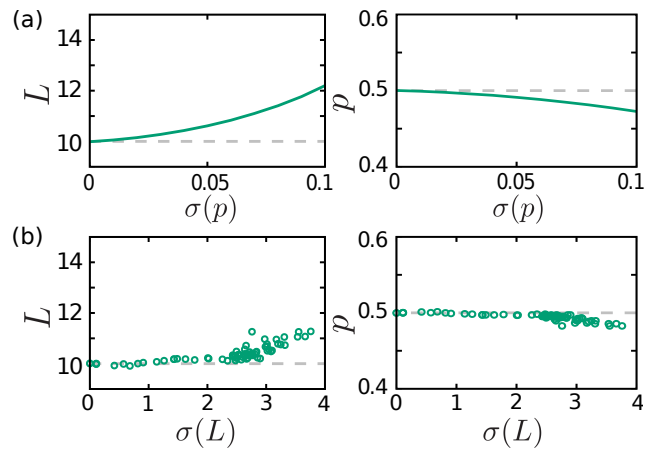}
\caption{The estimated values of the structural parameters $L$ and 
$p$ versus the structural disorder characterized by the variance of 
the parameter (a) $p$ and (b) $L$. The horizontal dashed lines indicate 
the actual parameter values. The parameter values (unless varied) 
are taken to be $L{=}10$ and $p{=}m{=}q{=}\frac12$. Each data point 
in panel (b) represents the result for a given static tree with a 
stochastic irregular branching pattern.}
\label{Fig7}
\end{figure}

\section{Disordered networks and experimental uncertainties}
\label{Sec:Disorder}

In derivation of the MFPT in the previous section, we considered 
a regular tree with the same tree extent along all branches and 
the same effective tendency to jump toward the target at every 
node. However, the real or virtual tree structures of interest 
are disordered. Particularly, heterogeneity in the degree of the 
nodes or in the capacity of the links has been widely studied in 
the context of complex networks. If the degree of the nodes or the 
weight of the links follow probability distributions, the effective 
tendency to move toward the target varies from node to node. The 
question arises of how far our analytical predictions for a regular 
tree remain valid when the network connectivity pattern becomes 
more and more diverse. To answer this question, we compare the 
analytical result of the MFPT for the constant $p$ across the tree, 
via Eq.\,(\ref{Eq:MFPT}), with the Monte Carlo simulation results 
where $p$ dynamically fluctuates at each node around the mean 
global value. Before each jump, we assign a new value to the 
effective tendency to move toward the target, which is randomly 
taken from a uniform probability distribution in the interval 
$[p{-}\delta p, p{+}\delta p]$.

Fig.\,\ref{Fig6}(a) shows that the deviations of the MFPT from 
the analytical expression (\ref{Eq:MFPT}) grow with increasing 
the variation range $\delta p$. It can be seen that the significance 
of the impact on the MFPTs depends on the mean value of $p$. The 
higher the tendency to jump toward the target node, the more robust 
the analytical predictions become. For comparison, we define an 
upper threshold of $10\%$ deviations of the MFPT from the theory 
and measure the critical value $\delta p_c$ for the variation 
range of the parameter $p$ in simulations, at which the fluctuations 
of the MFPT reach the $10\%$ threshold. As shown in Fig.\,\ref{Fig6}(b), 
broader fluctuations in $p$ can be tolerated with increasing $p$. 
For example, even a broad range of $15\%$ variations of $p$ around 
the mean value $p{=}0.7$ does not cause $10\%$ deviations 
in the measured MFPT from the predicted value. In regular trees 
with node degree $k{\gg}1$, $p{\rightarrow}0$ and the analytical 
predictions are only applicable in the limit of small variations 
in the degree of the nodes across the tree. As mentioned above, we 
have chosen a uniform distribution for $p$. In order to clarify 
whether the form of the probability distribution influences the 
MFPT results, we compare uniform and normal distributions of $p$ 
with the same mean and variance in Fig.\,\ref{Fig6}(c). The mean 
value is fixed at $\langle p \rangle{=}\frac12$ and a few examples 
for $\delta p$ have been examined. It can be seen that the 
differences between the FPT distributions are negligible when 
the first two moments of the distributions of the parameter $p$ 
are equal.

Next we investigate the influence of the structural disorder on the 
estimation of $L$ and $p$, as the main objective of the present study. 
We perform simulations where either $p$ or $L$ is allowed to fluctuate 
around its mean value. Let us first consider a tree with $L{=}10$ and 
allow $p$ to dynamically vary around $\langle p \rangle{=}\frac12$ in 
Monte Carlo simulations through the procedure which was previously 
explained in this section. By obtaining the MFPT from simulations 
and inserting it in Eq.\,(5), one can predict the structural parameters 
for a given standard deviation $\sigma(p)$. Assuming that one of the 
structural parameters $L$ or $p$ is known, we estimate the other one 
via Eq.\,(5) and compare it to its actual value in Fig.\,\ref{Fig7}(a). 
The deviation from the analytical prediction grows with increasing 
$\sigma(p)$, however, the effect is less pronounced for estimation 
of $p$ compared to $L$. We also note that the deviations are smaller 
for higher values of $\langle p \rangle$ (not shown). Our theoretical 
approach is thus applicable to treelike networks with moderate 
disorder in their connectivity pattern. We remind that according to 
Eq.\,(\ref{Eq:MFPT}) the MFPT approaches a linear scaling when 
$p{\rightarrow}1$, while it grows almost exponentially when $p$ 
decreases toward zero. Therefore, it is expected that the variation 
of $p$ around a mean value in the linear regime ($\langle p \rangle{>}
\frac12$) induces less deviations in the measured MFPTs, leading to 
a more accurate estimation of the structural parameters. On the other 
hand, fluctuations of $p$ around a mean value in the exponential regime 
($\langle p \rangle{<}\frac12$) causes an overestimation of the MFPT 
on average, which increases the errors of the estimated structural 
parameters. 

In order to study disorder in the extent of the tree, we generate 
static irregular trees with $\langle L \rangle{=}10$ and standard 
deviation $\sigma(L)$ in the following way: we randomly allow the 
nodes to have their child nodes in a hierarchical manner starting 
from the root node. The procedure continues until the average extent 
of the tree reaches $\langle L \rangle{=}10$. In Fig.\,7(b), we have 
presented the results for a tree with $p{=}\frac12$ and for $m{=}q
{=}\frac12$ as an example. The deviations from analytical predictions 
remain below $10\,\%$ even in considerably heterogeneous trees with 
$\frac{\sigma(L)}{\langle L \rangle}{\approx}40\,\%$.

\begin{figure}[t]
\centering
\includegraphics[width=0.47\textwidth]{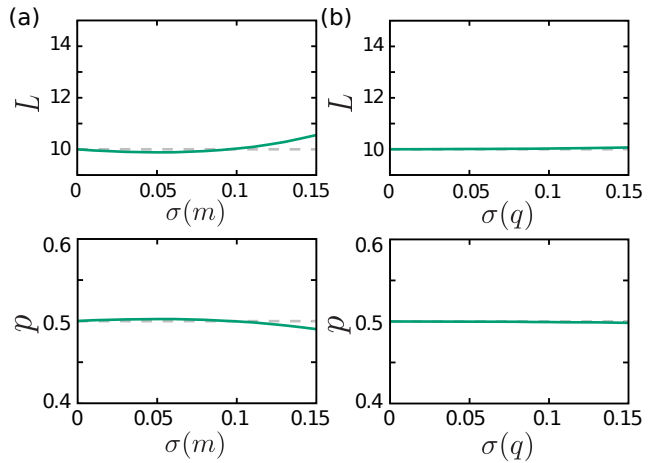}
\caption{The estimated values of the structural parameters $L$ and 
$p$ versus the experimental uncertainty characterized by the variance 
of the parameter (a) $m$ and (b) $q$. The horizontal dashed lines 
indicate the actual parameter values. The parameter values (unless 
varied) are taken to be $L{=}10$ and $p{=}m{=}q{=}\frac12$. The mean 
values of the variable parameters are $\langle m \rangle{=}\frac12$ 
and $\langle q \rangle{=}\frac12$ in panels (a) and (b), respectively.}
\label{Fig8}
\end{figure}

In addition to disorder in the structure of the network, one may introduce 
uncertainty in tunning the experimental parameters $m$ and $q$. For example, 
intermittent random walks induced by random trapping and release in cages 
or by stochastic temporal absorption along the path lead to statistical 
errors for the resulting waiting probabilities $m$ and $q$. In order to 
assess the robustness of our analytical estimation of $L$ and $p$ in the 
presence of experimental uncertainties, we perform simulations where the 
structure is regular but the parameters $m$ and $q$ fluctuate around their 
mean values. In each of the Monte Carlo simulations, we vary only one of 
the experimental parameters $m$ or $q$ while the other one is fixed at 
its mean value. A new value is assigned to the variable parameter at each 
random walk step, which is randomly taken from a uniform distribution with 
standard deviation $\sigma(m)$ or $\sigma(q)$. For a given fluctuation range 
[either $\sigma(m)$ or $\sigma(q)$], we obtain the MFPT from simulations. 
Next, we insert the resulting MFPT (as well as $\langle m \rangle$, $\langle 
q \rangle$, and one of the structural parameters $L$ or $p$) in 
Eq.\,(\ref{Eq:MFPT}) to predict the remaining structural parameter. In 
Fig.\,8, we compare the analytical estimations with the actual values to 
demonstrate the impact of uncertainty in the experimental parameters on 
the prediction of $L$ and $p$. It can be seen that the predicted values 
of $p$ or $L$ insignificantly deviate from the actual values for moderate 
(even up to $15\%$) experimental uncertainties. 

\section{Discussion and conclusion}
\label{Sec:Conclusion}

We presented an analytical framework to calculate the first-passage properties 
of intermittent random walks on treelike networks in terms of the waiting 
probabilities at the bulk nodes and leaves as well as the structural properties 
of the network such as its connectivity and extent. We proposed simple 
measurements of the mean-first-passage times of tracer particles entering 
the tree at the leaves and exiting from the root, to uniquely determine 
the structural properties. In regular trees, having access to the MFPT 
for only two different sets of waiting probabilities would be enough to 
unravel the structure. The idea of tuning the waiting probabilities of the 
random walkers to unravel the structure can be extended to other intrinsic 
dynamic properties such as the persistency of the walker \cite{Sadjadi15,Newby09}. 
Persistent random walkers can be employed to explore the structure of 
1D domains and obtain their length and the effective bias to hop toward the target, 
even though assigning a persistency to the particle dynamics on a network 
topology is not well defined in general. Calculating the higher moments of 
the FPT distribution is another possibility. Since the $i$-th moment $\langle
\tau^i\rangle = (z \frac{d}{d z})^iF(z)\Bigr|_{z{\rightarrow}1}$ can be 
straightforwardly obtained, having access to the first two moments of the 
FPT distribution also enables one to uniquely determine the structure via 
our analytical framework. However, the resulting set of equations may 
be redundant in special cases and it is considerably more difficult to 
evaluate variance or higher moments of the FPT distribution in experiments.

We addressed the validity range of the analytical predictions in networks 
with diverse connectivity patterns or heterogeneous branching morphologies. 
In the presence of disorder in the structural parameters $L$ and $p$ or 
uncertainty in the experimental parameters $m$ and $q$, even more 
measurements are required to determine the MFPT itself with a given 
accuracy. Therefore, it is naturally expected that the evaluation of 
the second moment of the FPT distribution with a given accuracy would 
be extremely time consuming in such cases. The advantage of considering 
intermittent random walks is that equipping particles with different 
diffusivity in experiments is possible, which makes our proposed 
measurements feasible in practice. The final remark is that the results 
obtained in this study are equivalently applicable to ordinary random 
walks on treelike structures which induce temporal absorption along 
the path or at the dead ends. A particularly relevant example is the 
transport in neuronal dendrites in the presence of biochemical cages 
along the dendritic tube.

\acknowledgments
We thank Z.\ Sadjadi and J.\ Kert\'esz for fruitful discussions. This work 
was funded by the Deutsche Forschungsgemeinschaft (DFG) through Collaborative 
Research Center SFB 1027 (Projects A7 and A8).

\end{document}